\documentclass[runningheads,envcountsame]{llncs}

\makeatletter \RequirePackage[bookmarks,unicode,colorlinks=true]{hyperref}%
\def\@citecolor{blue}%
\def\@urlcolor{blue}%
\def\@linkcolor{blue}%

\def\orcidID#1{\href{http://orcid.org/#1}{\smash{\protect\raisebox{-1.25pt}{\protect\includegraphics{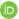}}}}}
\makeatother

\usepackage[dvipsnames]{xcolor}
\usepackage{hyperref}
\usepackage{mathtools}
\usepackage{multirow}
\usepackage{stmaryrd}
\usepackage{cancel}
\usepackage{amsmath,amssymb}
\usepackage{thmtools, thm-restate}
\usepackage{nccmath}
\usepackage[shortlabels]{enumitem}
\usepackage{microtype}
\usepackage{pbox}
\usepackage{marvosym}
\usepackage[noline,noend,linesnumbered]{algorithm2e}
\usepackage[hyperref=true, backend=bibtex, firstinits=true, maxbibnames=99, sortcites, style=numeric-comp]{biblatex}
\hypersetup{
	pdftitle={Control-Flow Refinement for Probabilistic Programs in KoAT}, colorlinks=true, linkcolor=blue, citecolor=olive, filecolor=magenta, urlcolor=cyan
}
\usepackage{todonotes}
\usepackage{placeins}
\usepackage{tikz}
\usepackage{caption}
\usepackage{subcaption}
\usepackage{mymatrix}
\usepackage{nicefrac}
\usepackage{wrapfig}

\usetikzlibrary{shapes,calc,arrows,automata,decorations.pathmorphing}
\usepackage[capitalize,nameinlink]{cleveref}
\usepackage{IEEEtrantools}

\newenvironment{myproof}{
	\noindent{\it Proof.}
}{\qed
	\medskip
}

\newcommand{\paper}[1]{}
\newcommand{\report}[1]{#1}
\newcommand{\final}[1]{}

\paper{\usepackage[hyperref=true, backend=bibtex, firstinits=true, maxbibnames=99, sortcites, style=numeric-comp]{biblatex}}

\usepackage[location=appendix,prependtoappendix=false,hideproofs=false]{moveproofs}
\makeatletter \newcommand*\bigcdot{\mathpalette\bigcdot@{.4}}
\newcommand*\bigcdot@[2]{\mathbin{\vcenter{\hbox{\scalebox{#2}{$\m@th#1\bullet$}}}}}
\makeatother

\renewcommand{\emptyset}{\varnothing}

\newcommand{\tool}[1]{\textsf{#1}}
\newcommand{\emphit}[1]{\textit{#1}}

\newcommand{\dontprint}[1]{}

\newcommand{\eat}[1]{}
\DeclarePairedDelimiter\abs{\lvert}{\rvert}

\newcommand{\NN}{\mathbb{N}}
\newcommand{\ZZ}{\mathbb{Z}}

\newcommand{\NNC}{\overline{\mathbb{N}}}

\newcommand{\RR}{\mathbb{R}}

\newcommand{\var}{\normalfont\texttt}

\newcommand{\true}{\var{true}}

\newcommand{\location}{\ell}
\newcommand{\state}{\sigma}
\newcommand{\initial}{\sigma_0}
\newcommand{\State}{\Sigma}
\newcommand{\update}{\eta}

\newcommand{\guard}{\varphi}

\newcommand{\landau}{\mathcal{O}}

\newcommand{\UTime}{{\mathcal{RB}}}

\newcommand{\BoundSet}{\mathcal{B}}

\newcommand{\Program}{\mathcal{P}}
\newcommand{\AtomSet}{\mathcal{A}}
\newcommand{\ConstraintSet}{\mathcal{C}}

\newcommand{\TSet}{\mathcal{T}}
\newcommand{\Set}{\mathcal{S}}
\newcommand{\GTSet}{\mathcal{GT}}
\newcommand{\VSet}{\mathcal{V}}
\newcommand{\PVSet}{\mathcal{PV}}

\newcommand{\SSet}{\mathcal{S}}
\newcommand{\LSet}{\mathcal{L}}

\newcommand{\IntProgram}{(\PVSet,\LSet,\location_0,\GTSet)}

\newcommand{\braced}[1]{\lbrace #1 \rbrace}

\crefname{definition}{Def.}{Def.}
\crefname{example}{Ex.}{Ex.}
\crefname{counterexample}{Counterex.}{Counterex.}
\crefname{appendix}{App.}{App.}
\crefname{ex}{Ex.}{Ex.}
\crefname{theorem}{Thm.}{Thm.}
\crefname{lemma}{Lemma}{Lemmas}
\crefname{remark}{Rem.}{Rem.}
\crefname{section}{Sect.}{Sect.}
\crefname{subsection}{Sect.}{Sect.}
\crefname{subsubsection}{Sect.}{Sect.}
\crefname{line}{Line}{Lines}
\crefname{corollary}{Cor.}{Cor.}
\crefname{figure}{Fig.}{Fig.}
\crefname{enumi}{}{}
\crefname{algorithm}{Alg.}{Alg.}

\newcommand{\KoAT}[0]{\tool{KoAT}}

\newcommand{\scheduler}{\mathfrak{S}}
\newcommand{\Scheduler}{\Pi}
\newcommand{\schedulerHistory}{\mathfrak{R}}
\newcommand{\SchedulerHistory}{\Pi^{\textsf{HD}}}
\newcommand{\prob}[2]{pr_{#1,\initial}}
\newcommand{\probTr}[2]{pr_{#1}}
\newcommand{\probP}[2]{pr_{#1,\initial}^{#2}}
\newcommand{\probM}[2]{\mathbb{P}_{#1,\initial}}
\newcommand{\probMP}[2]{\mathbb{P}^{#2}_{#1,\initial}}
\newcommand{\expP}[2]{\mathbb{E}_{#1,\initial}}
\newcommand{\expPP}[2]{\mathbb{E}_{#1,\initial}^{#2}}
\newcommand{\RandVar}[1]{\mathcal{R}}
\newcommand{\RandVarP}[1]{\mathcal{R}^{#1}}
\newcommand{\expRuntime}{\mathcal{R}_{\scheduler,\initial}}
\newcommand{\varexpRuntime}{\mathcal{R}_{\scheduler',\initial}}
\newcommand{\Run}{\mathsf{Run}}
\newcommand{\Conf}{\mathsf{Conf}}
\newcommand{\Path}{\mathsf{Path}}
\newcommand{\AdmPath}{\mathsf{AdmPath}}
\DeclareMathOperator{\embedding}{\beta}
\newcommand{\identity}{{\normalfont\textsf{id}}}

\title{Control-Flow Refinement for Complexity Analysis of Probabilistic Programs in \tool{KoAT}\thanks{funded by the Deutsche Forschungsgemeinschaft (DFG, German Research Foundation) - 235950644 (Project GI 274/6-2) and DFG Research Training Group 2236 UnRAVeL}}
\titlerunning{Control-Flow Refinement for Probabilistic Programs in \tool{KoAT}}
\author{Nils
  Lommen\final{$^{(\href{mailto:lommen@cs.rwth-aachen.de}{\mbox{\Letter}})}$}\orcidID{0000-0003-3187-9217}
  \and
  Éléanore Meyer\final{$^{(\href{mailto:eleanore.meyer@cs.rwth-aachen.de}{\mbox{\Letter}})}$}\orcidID{0000-0003-1038-4944}
  \and Jürgen Giesl\final{$^{(\href{mailto:giesl@cs.rwth-aachen.de}{\mbox{\Letter}})}$}\orcidID{0000-0003-0283-8520}}
\authorrunning{N.\
	Lommen et al.}
\institute{LuFG Informatik 2, RWTH Aachen University, Aachen, Germany\\
\email{\{lommen,eleanore.meyer,giesl\}@cs.rwth-aachen.de}}

\begin{document}
\allowdisplaybreaks \maketitle

\begin{abstract}
	Recently, we showed how to use control-flow refinement (CFR) to improve automatic complexity analysis of integer programs.
	While up to now CFR was limited to classical programs, in this paper we extend CFR to \emph{probabilistic} programs and show its soundness for complexity analysis.
	To demonstrate its benefits, we implemented our new CFR technique in our complexity analysis tool \tool{KoAT}.
\end{abstract}

\section{Introduction}
There exist numerous tools for complexity analysis of (non-probabilistic) programs, e.g., \cite{albert2008AutomaticInferenceUpper,albert2012CostAnalysisObjectoriented,albert2019ResourceAnalysisDriven,alias2010MultidimensionalRankingsProgram,avanzini2013CombinationFrameworkComplexity,carbonneaux2015CompositionalCertifiedResource,flores-montoya2016UpperLowerAmortized,frohn2017ComplexityAnalysisJava,giesl2022ImprovingAutomaticComplexity,hoffmann2017AutomaticResourceBound,lommen2023TargetingCompletenessUsing,lommen2022AutomaticComplexityAnalysis,moser2018JinjaBytecodeTerm,noschinski2013AnalyzingInnermostRuntime,sinn2017ComplexityResourceBound,brockschmidt2016AnalyzingRuntimeSize}.
Our tool \KoAT{} infers upper runtime and size bounds for (non-probabilistic) integer programs in a modular way\linebreak[2]
by analyzing subprograms separately and lifting the obtained results to global bounds on the whole program \cite{brockschmidt2016AnalyzingRuntimeSize}.
Recently, we developed several improvements of \KoAT{} \cite{giesl2022ImprovingAutomaticComplexity,lommen2022AutomaticComplexityAnalysis,lommen2023TargetingCompletenessUsing}
and showed that incorporating control-flow refinement (CFR) \cite{domenech2018IRankFinder,domenech2019ControlFlowRefinementPartial}
increases the power of automated complexity analysis significantly \cite{giesl2022ImprovingAutomaticComplexity}.

There are also several approaches for complexity analysis of \emph{probabilistic}
programs, e.g., \cite{avanzini2020ModularCostAnalysis,meyer2021InferringExpectedRuntimes,ngo2018BoundedExpectationsResource,schroer2023DeductiveVerificationInfrastructure,wang2020RaisingExpectationsAutomating,kaminski2018weakest,lexrsm,FoundationsExpectedRuntime2020,LeutgebCAV2022amor,KatoenPOPL23}.
In particular, we also adapted \KoAT{}'s approach for runtime and size bounds, and introduced a modular framework for automated complexity analysis of probabilistic integer programs in \cite{meyer2021InferringExpectedRuntimes}.
However, the improvements of \KoAT{} from \cite{giesl2022ImprovingAutomaticComplexity,lommen2022AutomaticComplexityAnalysis,lommen2023TargetingCompletenessUsing}
had not yet been adapted to the probabilistic setting.
In particular, we are not aware of any existing technique to combine CFR with complexity analysis of probabilistic programs.

Thus, in this paper, we develop a novel CFR technique for probabilistic pro\-grams which
could be used as a black box by every complexity analysis tool.
More\-over, to reduce the overhead by CFR, we integrated CFR natively into \KoAT{} by\linebreak[2]
calling it on-demand in a modular way.
Our experiments show that CFR increases the power of \tool{KoAT} for complexity analysis of probabilistic programs substantially.

The idea of CFR is to gain information on the values of program variables and to sort out infeasible program paths.
For example, consider the probabilistic \textbf{while}-loop \eqref{exa:while_original}.
Here, we flip a (fair) coin and either set $x$ to $0$ or do nothing.
\begin{equation}
	\textbf{while $\;x > 0\;$ do } \; x \gets 0 \; \oplus_{\nicefrac{1}{2}} \; \texttt{noop}
	\; \textbf{ end}\label{exa:while_original}
\end{equation}

\noindent
The update $x \gets 0$ is in a loop.
However, after setting $x$ to $0$,
the loop cannot be executed again.
To simplify its analysis, CFR ``unrolls'' the loop resulting in \eqref{exa:while_unrolled}.
\begin{align}
	 & \textbf{while $x > 0$ do } \; \texttt{break} \; \oplus_{\nicefrac{1}{2}} \;
	\texttt{noop} \; \textbf{ end}\nonumber                                  \\
	 & \textbf{if $x > 0$ then $x\gets 0$ end}\label{exa:while_unrolled}
\end{align}

\noindent
Here, $x$ is updated in a separate, \emph{non-probabilistic} \textbf{if}-statement and the loop does\linebreak[2]
not change variables.
Thus, we sorted out paths where $x \gets 0$ was executed repeatedly.
Now, techniques for probabilistic programs can be used for the \textbf{while}-loop.\linebreak[2]
The rest of the program can be analyzed by techniques for non-probabilistic programs.
In particular, this is important if \eqref{exa:while_original} is part of a larger program.

We present necessary preliminaries in \cref{sec:preliminaries}.
In \cref{sec:cfr}, we introduce our\linebreak[2]
new control-flow refinement technique and show how to combine it with automated complexity analysis of probabilistic programs.
We conclude in \cref{sec:conclusion} by an experi\-mental evaluation with our tool
\KoAT{}.
We refer to\report{ the appendix}\paper{ \cite{report}} for further details on probabilistic programs and the soundness proof of our CFR technique.

\section{Preliminaries}
\label{sec:preliminaries}
Let $\VSet$ be a set of variables.
An \emph{atom} is an inequation $p_1 < p_2$ for polynomials $p_1,p_2\in\ZZ[\VSet]$, and
the set of all atoms is denoted by $\AtomSet(\VSet)$.
A \emph{constraint} is a (possibly empty) conjunction of atoms, and
$\ConstraintSet(\VSet)$ denotes the set of all constraints.
In addition to ``$<$'', we also use ``$\geq$'', ``$=$'', etc., which can be simulated by constraints (e.g., $p_1 \geq p_2$ is equivalent to $p_2 < p_1 + 1$ for integers).

For \emph{probabilistic integer programs (PIPs)}, as in \cite{meyer2021InferringExpectedRuntimes}
we use a formalism based on transitions, which also allows us to represent \textbf{while}-programs like \eqref{exa:while_original} easily.
A PIP is a tuple $\IntProgram$ with a finite set of program variables $\PVSet\subseteq\VSet$, a finite set of locations $\LSet$, a fixed initial location $\location_0 \in \LSet$, and a finite set of general transitions $\GTSet$.
A \emph{general transition} $g\in\GTSet$ is a finite set of transitions which share the same start location $\ell_g$ and the same guard $\guard_g$.
A \emph{transition} is a 5-tuple $(\location,\guard,p,\update,\location')$ with a \emph{start location} $\location\in\LSet$, \emph{target location}
$\location'\in\LSet\setminus\braced{\location_0}$, \emph{guard} $\guard\in\ConstraintSet(\VSet)$, \emph{probability} $p\in [0,1]$, and \emph{update} $\update: \PVSet\to\ZZ[\VSet]$.
The probabilities of all transitions in a general transition add up to $1$.
We always require that general transitions are pairwise disjoint and let
$\TSet = \biguplus_{g\in\GTSet} g$ denote the set of all transitions.
PIPs may have \emphit{non-deterministic branching}, i.e., the guards of several transitions can be satisfied.
Moreover, we also allow \emphit{non-deterministic (temporary) variables} $\VSet\setminus\PVSet$.
To simplify the presentation, we do not consider transitions with individual costs and updates which use probability distributions, but the approach can easily be extended accordingly.
From now on, we fix a PIP $\Program = \IntProgram$.

\begin{example}
	The PIP in \Cref{fig:original_program} has $\PVSet = \{x,y\}$, $\LSet = \{\ell_0,
        \ell_1, \ell_2 \}$, and four general transitions $\{t_0\}$, $\{t_{1a}, t_{1b}\}$,
        $\{t_2\}$, $\{t_3\}$.	The transition $t_0$ starts at the initial location
        $\location_0$ and sets $x$ to a non-deterministic positive value $u \in
        \VSet\setminus \PVSet$, while $y$ is unchanged. (In \Cref{fig:original_program},
        we omitted unchanged updates like $\update(y) = y$,	the guard $\true$, and the
        probability $p = 1$ to ease readability.)
	If the general transition is a singleton, we often use transitions and general transitions interchangeably.
	Here, only $t_{1a}$ and $t_{1b}$ form a non-singleton general transition which corresponds to\linebreak[2]
	the program \eqref{exa:while_original}.
	We denoted such (probabilistic) transitions by dashed arrows in\linebreak[2]
	\Cref{fig:original_program}.
	We	extended \eqref{exa:while_original} by a loop of $t_2$ and $t_3$ which is
        only executed if $y > 0 \wedge x = 0$ (due to $t_2$'s guard) and decreases $y$ by $1$ in each iteration (via $t_3$'s update).
\end{example}

\begin{figure}[t]
	\centering
	\scalebox{0.8}{
		\begin{tikzpicture}[->,>=stealth',shorten >=1pt,auto,node distance=6cm,semithick,initial text=$ $]
			\node[state,initial left] (l0) {$\location_0$};
			\node[state] (l1) [right = 6.6cm of l0]{$\location_1$};
			\node[state] (l2) [right = 3.8cm of l1]{$\location_2$};
			\draw (l0) edge[above] node [align=center,xshift=-1.3cm] {$t_0 :\guard = (u > 0) \;\; \update(x) = u$} (l1);
			\draw (l1) edge [loop above,dashed,draw=purple] node
				[align=center,left,yshift=.125cm,xshift=0.2cm]
				{$\begin{array}{rcl}
							t_{1a}:\guard & = & (x > 0)         \\
							p             & = & \nicefrac{1}{2}\end{array}$} (l1);
			\draw (l1) edge [loop below,dashed,draw=purple] node
				[align=center,left,xshift=-.25cm]
				{$\begin{array}{rcl}
							t_{1b}:\guard & = & (x > 0)         \\
							\update(x)    & = & 0               \\
							p             & = & \nicefrac{1}{2}
						\end{array}$} (l1);
			\draw (l1) edge [bend right, below] node [align=center,below] {$t_2:\guard = (y> 0$\\\qquad\; $\wedge\; x = 0)$} (l2);
			\draw (l2) edge [bend right, above] node [align=center,above] {$t_3:\update(y) = y - 1$} (l1);
		\end{tikzpicture}
	}
	\vspace*{-.1cm}
	\caption{A Probabilistic Integer Program}\label{fig:original_program}
	\vspace*{-.4cm}
\end{figure}
A \emph{state} is a function $\state: \VSet\to\ZZ$, 
$\State$ denotes the set of all states, and a \emph{configuration} is a pair of a location
and a state. 
To extend finite sequences of configurations to infinite ones, we introduce a special location $\location_\bot$ (indicating termination) and a special transition $t_\bot$ (and its general transition $g_\bot = \braced{t_\bot}$) to reach the configurations of a run after termination.
Let $\LSet_\bot = \LSet\uplus\braced{\location_\bot}$, $\TSet_\bot =
\TSet\uplus\braced{t_\bot}$,  $\GTSet_\bot = \GTSet\uplus\braced{g_\bot}$, and let $\Conf
= (\LSet_\bot\times\State)$ denote the set of all configurations. 
A \emph{path} has the form $c_0\rightarrow_{t_1} \dots\rightarrow_{t_n} c_n$ for
$c_0,\dots,c_n\in\Conf$ and $t_1,\dots,t_n\in\TSet_\bot$ for an $n\in\NN$, and a
\emph{run} is an infinite path $c_0\rightarrow_{t_1} c_1 \rightarrow_{t_2} \cdots$. Let
$\Path$ and $\Run$ denote the sets of all paths and all runs, respectively.

We use Markovian schedulers $\scheduler: \Conf \to \GTSet_\bot\times\State$ to resolve all non-determinism.
For $c = (\ell,\sigma) \in \Conf$, a \emph{scheduler} $\scheduler$ yields a pair $\scheduler(c) = (g,\tilde{\sigma})$ where $g$ is the next general transition to be taken (with $\ell = \ell_g$) and $\tilde{\sigma}$ chooses values for the temporary variables such that $\tilde{\sigma} \models \guard_g$ and $\sigma(v) = \tilde{\sigma}(v)$ for all $v \in \PVSet$.
If $\GTSet$ contains no such $g$, we obtain $\scheduler(c) = (g_\bot,\sigma)$.
For the formal definition of Markovian schedulers, we refer to \report{\Cref{app:pip}}\paper{\cite{report}}.

For every $\scheduler$ and $\initial\in\State$, we define a probability mass function $\prob{\scheduler}{\Program}$.
For all\linebreak[2]
$c\in\Conf$, $\prob{\scheduler}{\Program}(c)$ is the probability that a run with scheduler $\scheduler$ and the initial\linebreak[2]
state $\sigma_0$ starts in $c$.
So $\prob{\scheduler}{\Program}(c) = 1$ if $c = (\location_0,\sigma_0)$ and $\prob{\scheduler}{\Program}(c) = 0$ otherwise.

For all $c, c' \in \Conf$ and $t \in \TSet_\bot$, let $\probTr{\scheduler}{\Program}(c \rightarrow_{t} c')$ be the probability that one goes
from $c$ to $c'$  via the transition $t$ when using the scheduler $\scheduler$  (see \report{\Cref{app:pip}}\paper{\cite{report}} for the formal definition of $\probTr{\scheduler}{\Program}$).
Then for any path $f = (c_0\rightarrow_{t_1} \dots\rightarrow_{t_n} c_n)\in\Path$, let $\prob{\scheduler}{\Program}(f) = \prob{\scheduler}{\Program}(c_0)\cdot \probTr{\scheduler}{\Program}(c_0 \rightarrow_{t_1} c_1)\cdot\ldots\cdot\probTr{\scheduler}{\Program}(c_{n-1}\rightarrow_{t_n}c_n)$.
Here, all paths $f$ which are not ``admissible'' (e.g., guards are not fulfilled,
transitions are starting or ending in wrong locations, etc.) have probability
$\prob{\scheduler}{\Program}(f) = 0$.

The semantics of PIPs can be defined via a corresponding probability space, obtained by a standard cylinder construction.
Let $\probM{\scheduler}{\Program}$ denote the probability measure which lifts $\prob{\scheduler}{\Program}$ to cylinder sets: For any $f\in\Path$, we have $\prob{\scheduler}{\Program}(f) = \probM{\scheduler}{\Program}(\text{Pre}_f)$ for the set $\text{Pre}_f$ of all infinite runs with prefix $f$.
So $\probM{\scheduler}{\Program}(\Theta)$ is the probability that a run from $\Theta \subseteq \Run$ is obtained when using the scheduler $\scheduler$ and starting in $\sigma_0$.
Let $\expP{\scheduler}{\Program}$ denote the associated expected value operator.
So for any random variable $X: \Run \to \NNC = \NN\cup\braced{\infty}$, we have $\expP{\scheduler}{\Program}(X) = \sum_{n\in\NNC}\; n\cdot\probM{\scheduler}{\Program}(X = n)$.
For a detailed construction, see\report{ \Cref{app:pip}}\paper{
  \cite{report}}.

\begin{definition}[Expected Runtime]
	For $g \in \GTSet$, $\RandVar{\Program}_g: \Run \to \NNC$ is a random variable
        with $\RandVar{\Program}_g(c_0\rightarrow_{t_1} c_1 \rightarrow_{t_2} \cdots) =
        |\braced{i\in\NN\mid t_i \in g}|$, i.e.,
 $\RandVar{\Program}_g(\vartheta)$ is the number of times that a transition from $g$ was applied in the run
        $\vartheta \in \Run$.
	Moreover, the random variable $\RandVar{\Program}: \Run \to \NNC$ denotes the number of transitions that were executed before termination, i.e., for all $\vartheta \in \Run$ we have $\RandVar{\Program}(\vartheta) = \sum_{g\in\GTSet} \RandVar{\Program}_g(\vartheta)$.
	For a scheduler $\scheduler$ and $\sigma_0 \in \Sigma$, the \emph{expected runtime} of $g$ is $\expP{\scheduler}{\Program}(\RandVar{\Program}_g)$ and the \emph{expected runtime} of the program is $\expRuntime = \expP{\scheduler}{\Program}(\RandVar{\Program})$.
\end{definition}

The goal of complexity analysis for a PIP is to compute a bound on its \emph{expected\linebreak[2]
	runtime complexity}.
The set of \emph{bounds} $\BoundSet$ consists of all functions from $\State\to\RR_{\geq 0}$.

\begin{definition}[Expected Runtime Bound and Complexity \cite{meyer2021InferringExpectedRuntimes}]
	\label{Expected Runtime Bound}
The function $\UTime:\GTSet\to\BoundSet$
	is an \emph{expected runtime bound}
	if $(\UTime(g))(\initial) \geq \sup_\scheduler \expP{\scheduler}{\Program}(\RandVar{g}_g)$ for all $\initial\in\State$ and all $g\in\GTSet$.
	Then $\sum_{g\in\GTSet} \UTime(g)$ is a bound on the \emph{expected runtime
        complexity} of the whole program, i.e., $\sum_{g\in\GTSet} ((\UTime(g))(\initial))\linebreak[2] \geq \sup_\scheduler \expRuntime$ for all $\initial\in\State$.
\end{definition}

\section{Control-Flow Refinement for PIPs}
\label{sec:cfr}
We now introduce our novel CFR algorithm for \emph{probabilistic} integer programs, based on the partial evaluation technique for non-probabilistic programs from \cite{domenech2018IRankFinder,domenech2019ControlFlowRefinementPartial,giesl2022ImprovingAutomaticComplexity}.
In particular, our algorithm coincides with the classical CFR technique when the program is non-probabilistic.
The goal of CFR is to transform a program $\Program$ into a program $\Program'$ which is ``easier'' to analyze.
\Cref{thm:soundness} shows the soundness of our approach, i.e., that $\Program$ and $\Program'$ have the same expected runtime complexity.

Our CFR technique considers ``abstract'' evaluations which operate on sets of states. These sets are characterized by conjunctions $\tau$ of constraints from $\ConstraintSet(\PVSet)$, i.e., $\tau$ stands for all states $\state\in\State$ with $\state\models\tau$.
We now label locations $\location$ by formulas $\tau$ which describe (a superset of) those
states $\state$ which can occur in
$\location$, i.e., where
a configuration
$(\location,\state)$ is reachable
from some initial configuration
$(\location_0,\initial)$.
We begin with labeling every location by the constraint $\true$.
Then we add new copies of the locations with refined labels $\tau$ by considering how the updates of transitions affect the constraints of their start locations and their guards.
The labeled locations become the new locations in the refined program.

Since a location might be reachable by different paths, we may construct several variants $\langle\location,\tau_1\rangle,\dots,\langle\location,\tau_n\rangle$ of the same original location $\location$.
Thus, the formulas $\tau$ are not necessarily invariants that hold for \emph{all} evaluations that reach a location $\ell$, but we perform a case analysis and split up a location $\ell$ according to the different sets of states that may reach $\ell$.
Our approach ensures that a labeled location $\langle \location,\tau \rangle$ can only be reached by configurations $(\location, \state)$ where $\state\models\tau$.

We apply CFR only \emph{on-demand} on a (sub)set of transitions $\Set\subseteq\TSet$ (thus, CFR can be performed in a \emph{modular} way for different subsets $\Set$).
In practice, we choose $\Set$ heuristically and use CFR only on transitions where our currently inferred runtime bounds are ``not yet good enough''.
Then, for $\Program = (\PVSet,\LSet,\location_0,\GTSet)$, the result of the CFR algorithm is the program $\Program' = (\PVSet,\LSet',\langle\location_0,\true\rangle,\GTSet')$ where $\LSet'$ and $\GTSet'$ are the smallest sets satisfying the properties \eqref{alg:initial}, \eqref{alg:abstract}, and \eqref{alg:transitions} below.

First, we require that for all $\location\in\LSet$, all ``original'' locations $\langle\location,\true\rangle$ are in $\LSet'$.
In these locations, we do not have any information on the possible states yet:
\begin{align}
	\label{alg:initial}
	\forall\; \location\in\LSet.\; \langle\location,\true\rangle\in\LSet'
\end{align}

\noindent
If we already introduced a location $\langle\location,\tau\rangle \in \LSet'$ and there is
a transition $(\location,\guard,p,\update,\location')\linebreak[2] \in\Set$, then \eqref{alg:abstract} requires that we also add the location $\langle\location',\tau_{\guard,\update,\ell'}\rangle$ to $\LSet'$.
The formula $\tau_{\guard,\update,\ell'}$ over-approximates the set of states that can result from states that satisfy $\tau$ and the guard $\guard$ of the transition when applying the update $\update$.
More precisely, $\tau_{\guard,\update,\ell'}$ has to satisfy $(\tau\wedge\guard)\models\update(\tau_{\guard,\update,\ell'})$.
For example, if $\tau = (x = 0)$, $\guard = \true$, and $\update(x) = x-1$, then we might have $\tau_{\guard,\update,\ell'} = (x=-1)$.

To ensure that every $\location'\in\LSet$
only gives rise to \emph{finitely} many new labeled locations $\langle \location', \tau_{\guard,\update,\ell'} \rangle$, we perform \emph{property-based abstraction}: For every location $\location'$, we use a finite so-called \emph{abstraction layer} $\alpha_{\location'} \subset \braced{p_1 \sim p_2\mid p_1,p_2\in\ZZ[\PVSet]\text{ and }\sim\;\in\braced{<,\leq,=}}$ (see \cite{domenech2019ControlFlowRefinementPartial} for heuristics to compute $\alpha_{\location'}$).
Then we require that $\tau_{\guard,\update,\ell'}$ must be a conjunction of constraints from $\alpha_{\location'}$ (i.e., $\tau_{\guard,\update,\ell'} \subseteq \alpha_{\location'}$ when regarding sets of constraints as their conjunction). This guarantees termination of our CFR algorithm, since for every location $\location'$ there are only finitely many possible labels.
\begin{align}
	\label{alg:abstract}
	\nonumber & \forall\; \langle\location,\tau\rangle\in\LSet'.\;\forall\; (\location,\guard,p,\update,\location')\in\Set.\; \langle\location',\tau_{\guard,\update,\ell'}\rangle\in\LSet' \\[-.1cm]
						& \hspace*{3cm} \text{where $\tau_{\guard,\update,\ell'} = \braced{\psi\in\alpha_{\location'}\mid(\tau\wedge\guard)\models\update(\psi)}$}
\end{align}

Finally, we have to ensure that $\GTSet'$ contains all ``necessary'' (general) transitions.
To this end, we consider all $g\in\GTSet$.
The transitions $(\location,\guard,p,\update,\location')$ in $g \cap \Set$ now have to connect the appropriately labeled locations.
Thus, for all labeled variants $\langle\location,\tau\rangle\in\LSet'$, we add the transition $(\langle\location,\tau\rangle,\tau\wedge\guard,p,\update,\langle\location', \tau_{\guard,\update,\ell'}\rangle)$.
In contrast, the transitions $(\location,\guard,p,\update,\location')$ in $g \setminus \Set$ only reach the location where $\location'$ is labeled with $\true$, i.e., here we add the transition $(\langle\location,\tau\rangle,\tau\wedge\guard,p,\update,\langle\location',\true\rangle)$.
\begin{align}
	\label{alg:transitions}
		& \forall\; \langle\location,\tau\rangle\in\LSet'.\;\forall\; g\in\GTSet.\nonumber                                                                                                                                  \\[-.1cm]
		& \left(\braced{(\langle\location,\tau\rangle,\tau\wedge\guard,p,\update,\langle\location', \tau_{\guard,\update,\ell'}\rangle) \mid (\location,\guard,p,\update,\location')\in g \cap \Set} \;\cup\right.\nonumber \\[-.1cm]
		& \left.\;\braced{(\langle\location,\tau\rangle,\tau\wedge\guard,p,\update,\langle\location',\true\rangle) \mid (\location,\guard,p,\update,\location')\in g \setminus \Set}\right)\qquad \in\GTSet'
\end{align}

$\LSet'$ and $\bigcup_{g\in\GTSet'}g$ are finite due to the property-based abstraction, as there are only finitely many possible labels for each location.
Hence, repeatedly ``unrolling'' transitions by \eqref{alg:transitions} leads to the (unique) least fixpoint.
Moreover, \eqref{alg:transitions} yields proper general tran\-sitions, i.e., their probabilities still add up to $1$.
In practice, we remove transitions\linebreak[2] with unsatisfiable guards, and locations that are not reachable from $\langle\location_0, \true\rangle$.\linebreak[2]
\Cref{thm:soundness} shows the soundness of our approach (see \report{\Cref{app:proofs}}\paper{\cite{report}} for its proof).

\begin{restatable}[Soundness of CFR for PIPs]{theorem}{soundness}
	\label{thm:soundness}
	\hspace*{-.2cm}	Let $\Program'\!=\!(\PVSet,\LSet',\langle\location_0,\true\rangle,\GTSet')$ be the PIP such that $\LSet'$ and $\GTSet'$ are the smallest sets satisfying \eqref{alg:initial}, \eqref{alg:abstract}, and \eqref{alg:transitions}.
	Let $\expRuntime^\Program$ and $\expRuntime^{\Program'}$ be the expected runtimes of $\Program$ and $\Program'$, respectively.
	Then for all $\sigma_0 \in \Sigma$ we have $\sup_{\scheduler}\expRuntime^\Program = \sup_{\scheduler}\expRuntime^{\Program'}$.
\end{restatable}
\makeproof{thm:soundness}{
	\soundness*

	\begin{myproof}
		Since	we now regard both the original program $\Program$ and the program $\Program'$ resulting from CFR, we add the considered program as a superscript when considering $\expRuntime$, $\expP{\scheduler}{\Program}$, $\RandVar{\Program}$, $\probM{\scheduler}{\Program}$, $\prob{\scheduler}{\Program}$, $\Path$, or $\Conf$.

		First, we prove $\sup_{\scheduler}\expRuntime^\Program \leq \sup_{\scheduler}\expRuntime^{\Program'}$ which is the crucial inequation for soundness.
		Afterwards, we complete the proof by showing $\sup_{\scheduler}\expRuntime^\Program \geq \sup_{\scheduler}\expRuntime^{\Program'}$.
		Thus, CFR does not increase the expected runtime complexity.

		As we will often restrict ourselves to the consideration of admissible paths only, we define the corresponding set $\AdmPath^{\Program}_{\scheduler,\initial} = \{f \in \Path^{\Program} \mid \probP{\scheduler}{\Program} (f) > 0\}$ of all such paths for a given program $\Program{}$, scheduler $\scheduler$, and initial state $\initial\in\State$.

		\paragraph*{\textbf{Soundness -- Proof of ``$\;\leq$'':}}
		Let $\initial\in\State$ be an initial state.
		We have to prove $\sup_{\scheduler}\expRuntime^\Program \leq \sup_{\scheduler}\expRuntime^{\Program'}$.
		In the following, we show that for every scheduler $\scheduler$ for $\Program$ there exists a scheduler $\scheduler'$ for $\Program'$ such that $\expRuntime^\Program \leq \varexpRuntime^{\Program'}$.
		Note that for any schedulers $\scheduler, \scheduler'$ and any $\initial \in \State$, we have
		\begin{align*}
			\expRuntime^\Program                                             & \leq \varexpRuntime^{\Program'}                                                                                                                                                \\
			\Leftrightarrow\expPP{\scheduler}{\Program}(\RandVarP{\Program}) & \leq \expPP{\scheduler'}{\Program'}(\RandVarP{\Program'}) \tag*{(definition of $\expRuntime^\Program$ and $\varexpRuntime^{\Program'}$)}
			\\
			\Leftrightarrow\sum_{n\in\NNC}n\cdot\probMP{\scheduler}{\Program}(\RandVarP{\Program}
			= n)                                                             & \leq \sum_{n\in\NNC}n\cdot\probMP{\scheduler'}{\Program'}(\RandVarP{\Program'} = n) \tag*{(definition of $\expPP{\scheduler}{\Program}$ and $\expPP{\scheduler'}{\Program'}$)}
		\end{align*}
		by definition.
		This can be equivalently expressed as
		\begin{align*}
			\sum_{n\in\NNC}n\cdot\probMP{\scheduler}{\Program}(\RandVarP{\Program}
			= n)         & \leq \sum_{n\in\NNC}n\cdot\probMP{\scheduler'}{\Program'}(\RandVarP{\Program'} = n)                                                 \\
			\Leftrightarrow\sum_{n\in\NNC_{> 0}}\probMP{\scheduler}{\Program}(\RandVarP{\Program}
			\geq n)      & \leq \sum_{n\in\NNC_{> 0}}\probMP{\scheduler'}{\Program'}(\RandVarP{\Program'} \geq n) \tag{$\NNC_{> 0} = \NNC\setminus\braced{0}$} \\
			\Leftrightarrow\sum_{n\in\NNC} \left(1 - \probMP{\scheduler}{\Program}(\RandVarP{\Program}
			< n) \right) & \leq \sum_{n\in\NNC} \left(1 - \probMP{\scheduler'}{\Program'}(\RandVarP{\Program'} < n) \right)
		\end{align*}
		since every run $\vartheta\in\Run$ with $\RandVarP{}(\vartheta) = n$
                occurs $n$-times in $\probMP{\scheduler}{}(\RandVarP{} \geq 1),
                \dots,\linebreak[2] \probMP{\scheduler}{}(\RandVarP{} \geq n)$.
		Now, we want to move from \emph{infinite} runs to \emph{finite} paths so that we can use an (injective) embedding which maps all paths of $\Program$ to paths of $\Program'$, i.e., to paths that use the transitions of $\GTSet'$ as constructed in \eqref{alg:transitions}.
		To this end, we define $\RandVarP{\Program}(c_0\to_{t_1}\dots\to_{t_n}c_n) = \abs{\braced{i\in\braced{1,\dots,n}\mid t_i\neq t_\bot}}$ for the path $c_0\to_{t_1}\dots\to_{t_n}c_n$ and the program $\Program$.
		We call a path $c_0\to_{t_1}\dots\to_{t_n}c_n$ \emph{terminated} if $t_1,\dots,t_{n-1} \neq t_\bot$ and $t_n = t_\bot$.
		Thus, we have
		\begin{align}
			\sum_{n\in\NNC}\left(1 - \probMP{\scheduler}{\Program}(\RandVarP{\Program} < n)\right) & \leq \sum_{n\in\NNC}\left(1 - \probMP{\scheduler'}{\Program'}(\RandVarP{\Program'} < n)\right) \nonumber     \\
			\Leftrightarrow\sum_{n\in\NNC} \left( 1 - \sum_{\substack{f\in\AdmPath^{\Program}_{\scheduler,\initial}                                                                     \\
					\RandVarP{\Program}(f) \,<\, n                                                                                                                                              \\
			\text{$f$ is terminated}}} \hspace*{-.2cm} \probP{\scheduler}{\Program}(f) \right)         & \leq \sum_{n\in\NNC} \left( 1 - \sum_{\substack{f\in\AdmPath^{\Program'}_{\scheduler',\initial} \\
					\RandVarP{\Program'}(f) \,<\, n                                                                                                                                             \\
					\text{$f$ is terminated}}} \hspace*{-.2cm} \probP{\scheduler'}{\Program'}(f) \right)\tag{$\dag$}\label{eq:proof_finite_paths}
			.
		\end{align}

		Next, we prove that for every scheduler $\scheduler: \Conf^{\Program}\to \GTSet_\bot\times\State$ there is a scheduler $\scheduler': \Conf^{\Program'}\to \GTSet'_\bot\times\State$ such that the last inequation \eqref{eq:proof_finite_paths} is valid.
		To this end, we use \cref{lem:embedding} which is presented below.
		For every $\initial \in \State$ and every scheduler $\scheduler: \Conf^{\Program}\to \GTSet_\bot\times\State$, \Cref{lem:embedding} yields a scheduler $\scheduler': \Conf^{\Program'}\to \GTSet'_\bot\times\State$ and a bijection $\embedding: \AdmPath^\Program_{\scheduler,\initial} \to \AdmPath^{\Program'}_{\scheduler',\initial}$ such that $\probP{\scheduler}{\Program}(f) = \probP{\scheduler'}{\Program'}(\embedding(f))$ for all $f\in\AdmPath^{\Program}_{\scheduler,\initial}$.
		Moreover, $\RandVar{\Program}(f) = \RandVar{\Program'}(\embedding(f))$ and $f$ is terminated iff $\embedding(f)$ is terminated.
		Hence, we have
		\begin{align*}
			\sum_{n\in\NNC} \left( 1 - \sum_{\substack{f\in\AdmPath^{\Program}_{\scheduler,\initial}                                                                                                             \\
			\RandVarP{\Program}(f) \,<\, n                                                                                                                                                                       \\
			\text{$f$ is terminated}}}\probP{\scheduler}{\Program}(f) \right) & = \sum_{n\in\NNC} \left( 1 - \sum_{\substack{f\in\embedding(\AdmPath^{\Program}_{\scheduler,\initial})                           \\
			\RandVarP{\Program'}(f) \,<\, n                                                                                                                                                                      \\
			\text{$f$ is terminated}}}\probP{\scheduler'}{\Program'}(\embedding(f)) \right)\tag{$\embedding$ is injective and $\probP{\scheduler}{\Program}(f) = \probP{\scheduler'}{\Program'}(\embedding(f))$} \\
			                                                                  & = \sum_{n\in\NNC} \left( 1 - \sum_{\substack{f\in\AdmPath^{\Program'}_{\scheduler',\initial}                                     \\
			\RandVarP{\Program'}(f) \,<\, n                                                                                                                                                                      \\
					\text{$f$ is terminated}}}\probP{\scheduler'}{\Program'}(f) \right) \tag{$\embedding$ is surjective, i.e., $\embedding(\AdmPath^{\Program}_{\scheduler,\initial}) = \AdmPath^{\Program'}_{\scheduler',\initial}$}
		\end{align*}
		which completes the proof of the soundness part ``$\leq$''.
		\begin{lemma}[Embedding of $\Path^{\Program}$ into $\Path^{\Program'}$]
			\label{lem:embedding}
			Let $\initial \in \State$ and $\scheduler: \Conf^\Program\to \GTSet_\bot\times\State$ be a scheduler.
			Then there exists a scheduler $\scheduler': \Conf^{\Program'}\to \GTSet'_\bot\times\State$ and a bijection $\embedding: \AdmPath^\Program_{\scheduler,\initial} \to \AdmPath^{\Program'}_{\scheduler',\initial}$ such that $\probP{\scheduler}{\Program}(f) = \probP{\scheduler}{\Program'}(\embedding(f))$ for all $f\in\AdmPath^{\Program}_{\scheduler,\initial}$.
			Moreover, $\RandVar{\Program}(f) = \RandVar{\Program'}(\embedding(f))$ and $f$ is terminated iff $\embedding(f)$ is terminated.
		\end{lemma}
		\begin{myproof}
			We define the scheduler $\scheduler'$ as follows:
			\begin{itemize}
				\item $\scheduler'(\langle\location,\tau\rangle,\state) =
                                  (g',\tilde{\state})$\quad if there exists a $g' \in
                                  \GTSet'$ with the start location
                                  $\langle\location,\tau\rangle$ where $\state
                                  \models\tau$ and $\scheduler(\location,\state) =
                                  (g,\tilde{\state})$, such that $g$ and $\langle\location,\tau\rangle$ yield $g'$ in \eqref{alg:transitions} (i.e., $g$ has the start location $\location$, $\guard_{g'}$ is $\tau \wedge \guard_g$, the transitions in $g$ and $g'$ have the same probabilities and updates, and the target locations of the transitions in $g'$ are labeled variants of the target locations of the transitions in $g$)
				\item $\scheduler'(\langle\location,\tau\rangle,\state) = (g_\bot,\state)$\quad otherwise
				\item $\scheduler'(\location_\bot,\state) = (g_\bot,\state)$\qquad for all $\state\in\State$
			\end{itemize}
			Clearly, $\scheduler'$ is a valid scheduler according to \cref{def:Scheduler}.

			Let $\initial\in\State$ be a state and $\scheduler, \scheduler'$ as above.
			We define the bijection $\embedding$ inductively:
			\paragraph*{Induction Base:}
			Let $f\in\AdmPath^\Program_{\scheduler,\initial}$ be an admissible path of length $0$, i.e., $f$ only consists of the configuration $c_0 = (\location_0,\initial)$ with $\probP{\scheduler}{\Program}(c_0) = 1$.
			We define $\embedding(c_0) = (\langle\location_0,\true\rangle,\initial)$.
			Thus, $\probP{\scheduler'}{\Program'}(\embedding(c_0)) = 1$.
			Otherwise, the path $\embedding(c_0)$ would not be admissible, as all admissible paths for program $\Program'$ of length $0$ must start in location $\langle\location_0,\true\rangle$ with state $\initial$.
			Hence, $\embedding(\braced{f\in\AdmPath^{\Program}_{\scheduler,\initial} \mid f \text{ has length $0$} }) = \braced{f\in\AdmPath^{\Program'}_{\scheduler',\initial} \mid f \text{ has length $0$} }$.
			Moreover, $\RandVar{\Program}(c_0) = \RandVar{\Program'}(\embedding(c_0)) = 0$ and both $c_0$ and $\embedding(c_0)$ are not terminated.

			\paragraph*{Induction Step:}
			Let $f\in\AdmPath^\Program_{\scheduler,\initial}$ be an admissible path of length $n + 1$ of the form $f = c_0\to_{t_1}\dots\to_{t_{n + 1}}c_{n+1}$ where $c_{n+1} = (\location_{n+1},\state_{n+1})$.
			By the induction hypothesis we have $$\probP{\scheduler}{\Program}(f_n) = \probP{\scheduler'}{\Program'}(\embedding(f_n))$$ where $f_n = (\location_0,\state_0)\to_{t_1}\dots\to_{t_n}(\location_n,\state_n)$ and $\embedding(f_n) = (\langle\location_0,\tau_0\rangle,\state_0)\to_{t_1'}\dots\to_{t_n'}(\langle\location_n,\tau_n\rangle,\state_n)$.

			If $t_{n} = t_{\bot}$, then the induction step holds by setting $\embedding (f) = (\embedding(f_{n}) \to_{t_{\bot}} (\location_{\bot},\state_{n}))$.
			In particular, then $\RandVar{\Program}(f) =\RandVar{\Program}(f_n)=\RandVar{\Program'}(\embedding(f_n)) = \RandVar{\Program'}(\embedding(f))$ by the induction hypothesis, and both $f$ and $\embedding(f)$ are terminated.

			Otherwise, by construction of \eqref{alg:initial} and
                        \eqref{alg:abstract}, there must be a location
                        $\langle\location_{n+1},\tau_{n+1}\rangle\linebreak[2] \in\LSet'$ such that $\state_{n+1}\models\tau_{n+1}$.
			Furthermore, there must also be a transition $t_{n+1}'$ whose start location is $\langle\location_n,\tau_n\rangle$ and a general transition of $\GTSet'$ which includes $t_{n+1}'$ by \eqref{alg:transitions}.
			In particular, $t_{n + 1}'$ and $t_{n+1}$ have the same probability and update, and the guard of $t_{n+1}'$ is the conjunction of $\tau_n$ and the guard of $t_{n + 1}$.
			Hence, $\probP{\scheduler}{\Program}(c_n\to_{t_{n + 1}}c_{n+1}) = \probP{\scheduler'}{\Program'}(c_n'\to (\langle\location_{n + 1},\tau_{n + 1}\rangle,\state_{n + 1}))$.
			Thus, defining $\embedding(f) = (\embedding(f_n)\to_{t_{n + 1}}(\langle\location_{n + 1},\tau_{n + 1}\rangle,\state_{n + 1}))$ yields the required properties.
			In particular, $\embedding$ is injective by the induction hypothesis for paths of length up to $n+1$.
			Moreover, $\RandVar{\Program}(f) =1 + \RandVar{\Program}(f_n)=1 + \RandVar{\Program'}(\embedding(f_n)) = \RandVar{\Program'}(\embedding(f))$ by the induction hypothesis, and both $f$ and $\embedding(f)$ are not terminated.

			Finally, we finish by proving that $\embedding$ is surjective, and hence bijective as well.
			Let $f' = (c_0\to_{t'_1}\dots\to_{t'_{n + 1}} c_{n + 1})\in\AdmPath^{\Program'}_{\scheduler',\initial}$ be an admissible path of length $n + 1$.
			By the induction hypothesis, there exists $f\in\AdmPath^{\Program}_{\scheduler,\initial}$ such that $\embedding(f) = (c_0\to_{t_1'}\dots\to_{t_{n}'} c_{n})$ holds.
			Since we fixed the scheduler $\scheduler$, there exists a unique configuration $(\location,\state)$ such that $f \to_{t_{n + 1}}
				(\location,\state)$ is admissible.
			Moreover, there is a unique $t'_{n + 1}$ in $\Program'$ that corresponds to $t_{n + 1}$ and whose start location is the location of $c_n$.
			Hence, we obtain $c_{n+1} =(\langle\location,\tau\rangle,\state)$ where $\langle\location,\tau\rangle$ is the target location of $t'_{n + 1}$.
			Thus, $\embedding(f\to_{t_{n + 1}} (\location,\state)) = f'$ which concludes the proof.
		\end{myproof}
		\paragraph*{\textbf{History-Dependent Schedulers:}}
		In the following tightness proof of $$\sup_{\scheduler}\expRuntime(\Program) \geq
                \sup_{\scheduler}\expRuntime(\Program'),$$
                we use \emph{history-dependent} schedulers.
		The reason for regarding history-dependent schedulers is that when removing the labels from the locations (in the step from $\Program'$ to $\Program$), different labeled locations $\langle\location, \tau_1\rangle$ and $\langle\location, \tau_2\rangle$ of $\Program'$ can be mapped to the same location $\location$ of $\Program$.
		Since the scheduler for $\Program'$ might behave differently on $\langle\location, \tau_1\rangle$ and $\langle\location, \tau_2\rangle$, this cannot be directly mimicked by a history-independent scheduler for $\Program$ which always has to behave in the same way for $\location$.
		However, this problem can be solved by regarding history-dependent schedulers that depend on the whole path up to the current configuration.
		For convenience, we always use $\scheduler$ when we consider schedulers as in \cref{def:Scheduler} and $\schedulerHistory$ if we refer to history-dependent schedulers as in the following definition.
		\begin{definition}[History-Dependent Scheduler]
			\label{def:SchedulerHistory}
			A function $\schedulerHistory: \Path \to \GTSet_\bot\times \State$ is a \emph{history-dependent scheduler} if for every $f = (\location_0,\state_0)\to^*(\location,\state) \in\Path$, $\schedulerHistory(f) = (g,\tilde{\state})$ implies:
			\begin{enumerate}[label=(\alph*)]
				\item \label{schedulerHistory:req1}
							$\state (x) = \tilde{\state} (x)$ for all $x\in\PVSet$.
				\item \label{schedulerHistory:req2}
							$\location$ is the start location $\location_g$ of the general transition $g$.
				\item \label{schedulerHistory:req3}
							$\tilde{\state}\models\guard_g$ for the guard $\guard_g$ of the general transition $g$.
				\item $g = g_{\bot}$ and $\tilde{\state} = \state$, if $\location = \location_\bot$ or no $g\in \GTSet,\tilde{\state}\in \State$ satisfy \cref{schedulerHistory:req1,schedulerHistory:req2,schedulerHistory:req3}.
			\end{enumerate}
			$\SchedulerHistory$ denotes the set of all history-dependent schedulers.
		\end{definition}

		As for history-independent schedulers, we also define $\prob{\schedulerHistory}{\Program}$ accordingly.
		For a path $(\location,\state)$ of length $0$, i.e., if the path only consists of a single configuration, we have $$\prob{\schedulerHistory}{\Program}(\location,\state) =
			\begin{cases}
				1 & \text{if $\location = \location_0$ and $\state = \initial$} \\
				0 & \, \text{otherwise.}
			\end{cases}
		$$ Furthermore, for a transition $t$, path $f = c_0\to_{t_1}\dots\to_{t_n}c_n$, configuration $c' = (\location_{c'},\state')$, and history-dependent scheduler $\schedulerHistory(f) = (g,\tilde{\sigma})$, we have $$\probTr{\schedulerHistory}{\Program}(f \to_t c') =
			\begin{cases}
				p & \text{if $t = (\location_c, \guard, p, \update, \location_{c'}) \in g$, $t \neq t_\bot$,} \\
					& \text{$\state'(v) = \tilde{\state}(\update(v))$ for all $v\in\PVSet$, and}                \\
					& \text{$\state'(v) = \tilde{\state}(v)$ for all $v\in\VSet\setminus\PVSet$
				}                                                                                             \\
				1 & \text{if $t = t_\bot \in g$, $\location_{c'} = \location_\bot$, and $\sigma' = \sigma$
				}                                                                                             \\
				0 & \, \text{otherwise.}
			\end{cases}
		$$ Then $\prob{\schedulerHistory}{\Program}(c_0\to_{t_1}\dots\to_{t_n}c_n) = \prob{\schedulerHistory}{\Program}(c_0)\cdot\prod_{i = 1}^n\probTr{\schedulerHistory}{\Program}(c_0\to_{t_1}\dots\to_{t_{i}} c_i)$.
		Analogous to the case for history-independent schedulers, for any $f \in \Path$, we say that $f$ is \emph{admissible} for $\schedulerHistory$ and $\sigma_0$ if $\prob{\schedulerHistory}{\Program}(f) > 0$.
		Similarly, $\probM{\schedulerHistory}{\Program}$ denotes the probability measure which lifts $\prob{\schedulerHistory}{\Program}$ to the sigma-algebra generated by all cylinder sets: For any path $f\in\Path$, we have $\prob{\schedulerHistory}{\Program}(f) = \probM{\schedulerHistory}{\Program}(\text{Pre}_f)$ for the set $\text{Pre}_f$ of all infinite runs with prefix $f$.
		$\expP{\schedulerHistory}{\Program}$ denotes the associated expected value operator.
		So for any random variable $X: \Run \to \NNC$, we have $\expP{\schedulerHistory}{\Program}(X) = \sum_{n\in\NNC}\; n\cdot\probM{\schedulerHistory}{\Program}(X = n)$.

	\paragraph*{\textbf{Tightness -- Proof of ``$\;\geq$'':}}
		We have to show that for every $\initial\in\State$ the inequation $$\sup_{\scheduler}\expRuntime(\Program) \geq \sup_{\scheduler}\expRuntime(\Program')$$ holds.
                To this end, we
		show that for every $\initial\in\State$
                we have $$\sup_{\scheduler\in\Scheduler}\expPP{\scheduler}{\Program'}(\RandVarP{\Program'}) \underset{(1)}{\leq}
			\sup_{\schedulerHistory\in\SchedulerHistory}\expPP{\schedulerHistory}{\Program'}(\RandVarP{\Program'}) \underset{(2)}{\leq}
			\sup_{\schedulerHistory\in\SchedulerHistory}\expPP{\schedulerHistory}{\Program}(\RandVarP{\Program}) \underset{(3)}{\leq}
			\sup_{\scheduler\in\Scheduler}\expPP{\scheduler}{\Program}(\RandVarP{\Program})$$
		\paragraph*{Proving (1):}
		Let $\scheduler: \Conf^{\Program'}\to\GTSet'_\bot\times\State$ be a scheduler and $\initial\in\State$.
		Defining the history-dependent scheduler $\schedulerHistory: \Path^{\Program'}\to\GTSet'_\bot\times\State$ by setting $\schedulerHistory(\dots\to c) = \scheduler(c)$ yields $\expPP{\scheduler}{\Program'}(\RandVarP{\Program'}) = \expPP{\schedulerHistory}{\Program'}(\RandVarP{\Program'})$.
		Hence, we have $\sup_{\scheduler \in \Scheduler}\expPP{\scheduler}{\Program'}(\RandVarP{\Program'}) \leq \sup_{\schedulerHistory\in\SchedulerHistory}\expPP{\schedulerHistory}{\Program'}(\RandVarP{\Program'})$ which proves the first inequation.
		\paragraph*{Proving (2):}
		Let $\schedulerHistory': \Path^{\Program'}\to\GTSet_\bot'\times\State$ be a history-dependent scheduler and $\initial\in\State$.
		We now have to define the history-dependent scheduler $\schedulerHistory: \Path^{\Program}\to\GTSet_\bot\times\State$.
		Let $f = (\location_0,\state_0)\to_{t_1}\dots\to_{t_n}(\location_n,\state_n)\in\Path^{\Program}$ be a path in $\Program$.
		If $f$ is not admissible, then we define $\schedulerHistory(f)$ arbitrarily such that \Cref{def:SchedulerHistory} is fulfilled.
		If $f$ is admissible, then there exists a unique corresponding path $f'\in\AdmPath^{\Program'}_{\schedulerHistory',\initial}$ with $f' = (\langle\location_0,\tau_0\rangle,\state_0)\to_{t_1'}\dots\to_{t_n'}(\langle\location_n,\tau_n\rangle,\state_n)$ (as in the surjectivity proof of $\beta$ in the proof of \Cref{lem:embedding}).
		Let $\schedulerHistory'(f') = (g',\state)$.
		Then we define $\schedulerHistory(f) = (g,\state)$ where $g'$ results from $g$ and \eqref{alg:transitions}.
		Our construction yields $\expPP{\schedulerHistory'}{\Program'}(\RandVarP{\Program'}) = \expPP{\schedulerHistory}{\Program}(\RandVarP{\Program})$.
		Thus, we have $\sup_{\schedulerHistory \in \SchedulerHistory}\expPP{\schedulerHistory}{\Program'}(\RandVarP{\Program'}) \leq \sup_{\schedulerHistory\in\SchedulerHistory}\expPP{\schedulerHistory}{\Program}(\RandVarP{\Program})$.
		\paragraph*{Proving (3):}
		Let $\schedulerHistory: \Path^\Program\to\GTSet_\bot\times\State$ be a history-dependent scheduler and $\initial\in\State$.
		We now consider the Markov decision process (MDP) $\mathcal{M}_{n}$ resulting from all admissible paths of the program $\Program$ under the scheduler $\schedulerHistory$ of length less than or equal to $n \in \NN$ starting in $c_0 = (\location_0,\initial)$.
		Formally, this MDP is defined as the tuple $\mathcal{M}_{n} = (\NN,S_{n},A_{n}, p, r)$ \cite[Section 2.1.3]{putermanMarkovDecisionProcesses} where:
		\begin{enumerate}
			\item $\NN$ is the set of decision epochs.
			\item $S_{n} \subseteq \Conf$ is the state set, where $S_n$ consists of all configurations $c_i$ with $i\in\braced{0,\dots,n}$ which occur on an admissible path $f = c_{0} \to_{t_{1}} \dots \to_{t_{i}}
				      c_{i} \in \AdmPath^{\Program}_{\schedulerHistory,\initial}$.
			\item For every $c \in S_n$, $A_{n}(c) \subseteq \AdmPath^{\Program}_{\schedulerHistory,\initial}$ is the set of actions in $c$, i.e., it is the smallest set of all admissible paths $f = c_{0} \to_{t_{1}} \dots \to_{t_{i}} c_i \in\AdmPath^{\Program}_{\schedulerHistory,\initial}$ with $i\in\braced{0,\dots,n - 1}$ and $c_{i} = c$.
			\item $p (c,f) (c') = \sum_{t \in g}pr_{\schedulerHistory}^{\Program}(f \to_{t} c')$ determines the probability of transitioning from state $c$ to state $c'$ when action $f \in A_{n}(c) \subseteq \AdmPath^{\Program}_{\schedulerHistory,\initial}$ is chosen, where $\schedulerHistory(f) = (g,\tilde{\sigma})$.
			\item $r(f)$ is the reward of the action $f$.
			      For an action $f \in \AdmPath^{\Program}_{\schedulerHistory,\initial}$, we have $r(f) = 0$ if $\schedulerHistory(f) = (g_{\bot},\tilde{\sigma})$ for some $\tilde{\sigma}$ and $r(f) = 1$, otherwise.
		\end{enumerate}
		Note that the MDP $\mathcal{M}_{n}$ might get ``stuck'', i.e., there might exist a reachable state $c = (\location,\state)\in S_{n}$ such that $A_{n}
			(c) = \emptyset$.
		However, this can be avoided by introducing a novel action $a_{c} \in A_{n}(c)$ leading to a dummy state $(\location_{\bot},\state)$ with reward $0$.
		In this dummy state, we introduce an additional rewardless action that only allows transitioning from $(\location_{\bot},\state)$ to $(\location_{\bot},\state)$ with probability $1$.

		The crucial observation is that the fixed scheduler $\schedulerHistory$ and the fixed initial configuration $c_0$ lead to only finitely many admissible paths of length at most $n$.
		Hence, all sets $S_{n}$ and $A_{n} (s)$ are finite.
		Thus by \cite[Theorem 7.1.9]{putermanMarkovDecisionProcesses}, there exists an optimal stationary and deterministic scheduler $\mathfrak{M}$ for $\mathcal{M}_{n}$ that maximizes the expected total reward.
		Now, we define the corresponding scheduler $\scheduler_n:\Conf^\Program\to\GTSet_\bot\times\State$ for the program $\Program$ by $\scheduler_{n}(c) = \schedulerHistory(\mathfrak{M}(c))$ for all $c\in S_n$.
		Otherwise, i.e., if $c\in \Conf\setminus S_n$, then $\scheduler_n(c)$ can be defined arbitrarily.

		In the following, let $\min(\RandVarP{\Program}, n)$ be the random variable with $\min(\RandVarP{\Program}, n)(\vartheta) = \min(\RandVarP{\Program}(\vartheta), n)$ for all $\vartheta\in\Run$.
		Now, by construction of the MDP $\mathcal{M}_{n}$ and optimality of the scheduler $\mathfrak{M}$, we have
		\begin{align*}
			\expPP{\scheduler_n}{\Program}(\RandVarP{\Program}) \geq \expPP{\schedulerHistory}{\Program}(\min(\RandVarP{\Program}, n)) & = \sum_{i\in\NNC}i\cdot\probMP{\schedulerHistory}{\Program}(\min(\RandVarP{\Program}, n) = i)                                        \\
			                                                                                                                           & = \sum_{i = 0}^n i\cdot\probMP{\schedulerHistory}{\Program}(\RandVarP{\Program} = i) \tag{$\ddag$}\label{proof:main_property_of_M_n}
		\end{align*}
		as we can fully model the first $n$ steps of $\Program$'s execution under scheduler $\schedulerHistory$ within the MDP $\mathcal{M}_{n}$.
		Thus, we have
		\begin{align}
			\sup_{\scheduler\in\Scheduler}\expPP{\scheduler}{\Program}(\RandVarP{\Program}) & \geq \lim_{n\to\infty}\expPP{\scheduler_n}{\Program}(\RandVarP{\Program})\tag{as $\scheduler_n\in\Scheduler$ for all $n\in\NN$}\nonumber                        \\
			                                                                                & \geq \lim_{n\to\infty} \sum_{i = 0}^n i\cdot\probMP{\schedulerHistory}{\Program}(\RandVarP{\Program} = i) \tag{by \eqref{proof:main_property_of_M_n}} \nonumber \\
			                                                                                & = \sum_{i\in\NNC}i\cdot \probMP{\schedulerHistory}{\Program}(\RandVarP{\Program} = i)                                                                           \\
			                                                                                & = \expPP{\schedulerHistory}{\Program}(\RandVarP{\Program}).\nonumber
		\end{align}
		Hence, $\sup_{\scheduler\in\Scheduler}\expPP{\scheduler}{\Program}(\RandVarP{\Program}) \geq \sup_{\schedulerHistory\in\SchedulerHistory}\expPP{\schedulerHistory}{\Program}(\RandVarP{\Program})$.

	\end{myproof}
}

\paragraph*{CFR Algorithm and its Runtime:}
To
implement the fixpoint construction of \Cref{thm:soundness} (i.e., to compute the PIP
$\Program'$),
our algorithm starts by introducing all ``original'' locations $\langle \location,\true\rangle$ for $\location\in\LSet$ according to \eqref{alg:initial}.
Then it iterates over all labeled locations $\langle\location,\tau\rangle$ and all
transitions $t \in \TSet$.
If the start location of $t$ is $\location$, then the algorithm extends
$\GTSet'$ by a
new transition
according to \eqref{alg:transitions}.
Moreover, it also adds the corresponding labeled target location to $\LSet'$ (as in
\eqref{alg:abstract}), if $\LSet'$ did not contain this labeled location yet.
Afterwards, we mark $\langle\location,\tau\rangle$ as finished and proceed with
a previously computed labeled location that is not marked yet.
So our implementation iteratively ``unrolls'' transitions by \eqref{alg:transitions} until
no new labeled locations are obtained (this yields the least fixpoint mentioned above).
Thus, unrolling steps with transitions from $\TSet\setminus\SSet$ do not invoke further computations.

To over-approximate the runtime of this algorithm, note that for every location $\location
\in \LSet$, there can be at most $2^{|\alpha_\location|}$ many labeled locations of the form $\langle\location,\tau\rangle$.
So if $\LSet = \braced{\location_0,\dots,\location_n}$, then the overall number of labeled locations is at most $2^{|\alpha_{\location_0}|} + \ldots + 2^{|\alpha_{\location_n}|}$.
Hence, the
algorithm performs at most $|\TSet| \cdot (2^{|\alpha_{\location_0}|} + \ldots +
2^{|\alpha_{\location_n}|})$
unrolling steps.

\begin{figure}[t]
	\centering \scalebox{0.8}{
		\begin{tikzpicture}[->,>=stealth',shorten >=1pt,auto,node distance=6cm,semithick,initial text=$ $]
			\node[state,initial left] (l0) {$\location_0$};
			\node[state] (l1) [right = 3.6cm of l0]{$\location_1$};
			\node[state] (l1a) [right = 3.3cm of l1]{\scriptsize$\langle \location_1,x=0\rangle$};
			\node[state] (l2) [right = 2.3cm of l1a]{\scriptsize$\langle \location_2,x=0\rangle$};
			\draw (l0) edge[below] node [align=center,xshift=-.3cm]
				{$\begin{array}{rcl}
							t_0': \guard & = & (u > 0) \\\update(x) &=& u
						\end{array}$} (l1);
			\draw (l1) edge [loop above,dashed,draw=purple] node
				[align=center,left] {$\begin{array}{rcl}
							t_{1a}':\guard & = & (x > 0)         \\
							p              & = & \nicefrac{1}{2}
						\end{array}$} (l1);
			\draw (l1) edge [dashed,draw=purple] node [align=center]
				{$\begin{array}{rcl}
							t_{1b}':\guard & = & (x > 0)         \\
							\update(x)     & = & 0               \\
							p              & = & \nicefrac{1}{2}
						\end{array}$} (l1a);
			\draw (l1a) edge [bend right, below] node [align=center,below] {$t_2':\guard = (y> 0$\\\qquad\; $\wedge\; x = 0)$} (l2);
			\draw (l2) edge [bend right, above] node
				[align=center,above] {$\begin{array}{rcl}
							t_3':\guard & = & (x=0) \\
							\update(y)  & = & y - 1
						\end{array}$} (l1a);
		\end{tikzpicture}
	}
	\vspace*{-.1cm}
	\captionof{figure}{Result of Control-Flow Refinement with $\Set = \braced{t_{1a},t_{1b},t_2,t_3}$}\label{fig:cfr_program}
	\vspace*{-.2cm}
\end{figure}

\begin{example}
	For the PIP in \Cref{fig:original_program} and $\Set = \braced{t_{1a},t_{1b},t_2,t_3}$,
	by \eqref{alg:initial}
	we start with $\LSet'=\{\langle\location_i,\true\rangle\mid i\in\braced{0,1,2}\}$.
	We abbreviate $\langle\location_i,\true\rangle$ by $\location_i$ in the final result of the CFR algorithm in \Cref{fig:cfr_program}.
	As $t_0 \in\braced{t_0}\setminus\Set$, by \eqref{alg:transitions}
	$t_0$ is redirected such that it starts at $\langle\location_0,\true\rangle$ and ends in $\langle\location_1,\true\rangle$, resulting in $t_0'$.
	We always use primes to indicate the correspondence between new and original transitions.

	Next, we consider $\braced{t_{1a},t_{1b}} \subseteq \Set$ with the guard $\guard = (x > 0)$ and start location $\langle \location_1, \true \rangle$.
	We first handle $t_{1a}$ which has the update $\update = \identity$.
	We use the abstraction layer $\alpha_{\location_0} = \emptyset$, $\alpha_{\location_1} = \braced{x = 0}$, and $\alpha_{\location_2} = \braced{x = 0}$.
	Thus,
	we have to find all $\psi \in \alpha_{\location_1} = \braced{x = 0}$ such that $(\true \wedge x>0) \models \update(\psi)$.
	Hence, $\tau_{x>0, \identity, \location_1}$ is the empty conjunction $\true$ as no $\psi$ from $\alpha_{\location_1}$ satisfies this property.
	We obtain
				\begin{align*}
				t_{1a}': & \; (\langle\location_1,\true\rangle,x > 0,\nicefrac{1}{2},\identity,\langle\location_1,\true\rangle).
			\end{align*}

	In contrast, $t_{1b}$ has the update $\update(x) = 0$.
	To determine $\tau_{x>0, \update, \location_1}$, again we have to find all $\psi\in \alpha_{\location_1} = \braced{x = 0}$ such that $(\true \wedge x>0) \models \update(\psi)$.
	Here, we get $\tau_{x>0, \update, \location_1} = (x = 0)$.
	Thus, by \eqref{alg:abstract} we create the location $\langle\location_1,x = 0\rangle$ and obtain
	\begin{align*}
				t_{1b}': & \; (\langle\location_1,\true\rangle,x > 0,\nicefrac{1}{2},\update(x) = 0,\langle\location_1,x = 0\rangle).
			\end{align*}

	\noindent
	As $t_{1a}$ and $t_{1b}$ form one general transition, by $\eqref{alg:transitions}$ we obtain $\{ t_{1a}', t_{1b}' \} \in \GTSet'$.

	Now, we consider transitions resulting from $\braced{t_{1a},t_{1b}}$ with the start location $\langle\location_1,x = 0\rangle$.
	However, $\tau = (x=0)$ and the guard $\guard = (x > 0)$ are conflicting, i.e., the transitions would have an unsatisfiable guard $\tau \wedge \guard$ and are thus omitted.

	Next, we consider transitions resulting from $t_2$ with $\langle\location_1,\true\rangle$ or $\langle\location_1,x = 0\rangle$ as their start location.
	Here, we obtain two (general) transitions $\{t_2'\}, \{t_2''\} \in \GTSet'$:
	\begin{align*}
		t_{2}': (\langle\location_1,x = 0  & \rangle, y > 0 \wedge x = 0,1,\identity,\langle\location_2,x = 0\rangle) \\[-.1cm]
		t_{2}'': (\langle\location_1,\true & \rangle, y > 0 \wedge x = 0,1,\identity,\langle\location_2,x=0\rangle)
	\end{align*}

	However, $t_2''$ can be ignored since $x = 0$ contradicts the invariant $x > 0$ at $\langle\location_1,\true\rangle$.
	\KoAT{} uses \tool{Apron} \cite{jeannet2009ApronLibraryNumerical} to infer invariants like $x > 0$ automatically.
	Finally, $t_3$ leads to the transition $t_{3}': (\langle\location_2,x = 0\rangle,x=0,1,\update(y) = y - 1,\langle\location_1,x = 0\rangle)$.
	Thus, we obtain $\LSet' = \braced{\langle\location_i,\true\rangle\mid i\in\braced{0,1}}\cup\braced{\langle\location_i,x=0\rangle\mid i\in\braced{1,2}}$.
\end{example}

\KoAT{} infers a bound $\UTime(g)$ for each $g \in \GTSet$ individually (thus, non-probabilis\-tic program parts can be analyzed by classical techniques).
Then $\sum_{g\in\GTSet}\UTime(g)$\linebreak[2]
is a bound on the expected runtime complexity of the
whole program, see \Cref{Expected Runtime Bound}.

\begin{example}
	We now infer a bound on the expected runtime complexity of the PIP in \Cref{fig:cfr_program}. Transition $t_0'$ is not on a cycle, i.e., it can be evaluated at most once.\linebreak[2]
	So $\UTime(\braced{t_0'}) = 1$ is an (expected) runtime bound for the general transition $\braced{t_0'}$.

	For the general transition $\braced{t_{1a}',t_{1a}'}$, \KoAT{} infers the expected
        runtime bound $2$\linebreak[2]
        via probabilistic linear ranking functions (PLRFs, see e.g., \cite{meyer2021InferringExpectedRuntimes}).
	More precisely, \KoAT{} finds the \emph{constant} PLRF $\braced{\location_1 \mapsto 2, \langle\location_1,x = 0\rangle \mapsto 0}$.
	In contrast, in the orig\-inal program of \Cref{fig:original_program}, $\braced{t_{1a}, t_{1b}}$ is not decreasing w.r.t.\ any constant PLRF, be\-cause $t_{1a}$ and $t_{1b}$ have the same target location. So here, every PLRF where $\braced{t_{1a},\linebreak[2]
			t_{1b}}$ decreases in expectation depends on $x$.
	However, such PLRFs do not yield a finite runtime bound in the end, as $t_0$ instantiates $x$ by the non-deterministic value $u$.
	Therefore, \KoAT{} fails on the program of \Cref{fig:original_program} without using CFR.

	For the program of \Cref{fig:cfr_program}, \KoAT{} infers $\UTime(\braced{t_2'}) =\UTime(\braced{t_3'}) =y$.
	By adding all runtime bounds, we obtain the bound $3+2\cdot y$ on the expected
        runtime complex\-ity of the program in \Cref{fig:cfr_program} and thus by
        \Cref{thm:soundness} also of the program in \Cref{fig:original_program}.
\end{example}

\section{Implementation, Evaluation, and Conclusion}
\label{sec:conclusion}

We presented a novel control-flow refinement technique for probabilistic programs and proved that it does not modify the program's expected runtime complexity.
This allows us to combine CFR with approaches for complexity analysis of probabi\-listic programs.
Compared to its variant for non-probabilistic programs, the sound\-ness proof of \Cref{thm:soundness} for probabilistic programs is considerably more involved.

Up to now, our complexity analyzer \KoAT{} used the tool \tool{iRankFinder} \cite{domenech2018IRankFinder} for CFR of non-probabilistic programs \cite{giesl2022ImprovingAutomaticComplexity}.
To demonstrate the benefits of CFR for complexity analysis of probabilistic programs, we now replaced the call to \tool{iRankFinder} in \KoAT{} by a native implementation of our new CFR algorithm.
\KoAT{}\linebreak[2]
is written in \tool{OCaml} and it uses \tool{Z3} \cite{demoura2008Z3EfficientSMT} for SMT solving, \tool{Apron} \cite{jeannet2009ApronLibraryNumerical} to generate invariants, and the \tool{Parma Polyhedra Library} \cite{bagnara2008ParmaPolyhedraLibrarya} for computations with polyhedra.

We used all 75 probabilistic benchmarks from \cite{ngo2018BoundedExpectationsResource,meyer2021InferringExpectedRuntimes} and added 15 new benchmarks including our leading example and problems adapted from the \emph{Termination Problem Data Base} \cite{tpdb}, e.g., a probabilistic version of McCarthy's 91 function.
Our benchmarks also contain examples where CFR is useful even if it cannot sepa\-rate probabilistic from non-probabilistic program parts as in our leading example.

\Cref{fig:evaluation} shows the results of our experiments.
We compared the configuration of \KoAT{} with CFR (``\tool{KoAT\,\!+\,\!CFR}'') against \KoAT{} without CFR.
Moreover, as in \cite{meyer2021InferringExpectedRuntimes}, we also compared with the main other recent tools for inferring upper bounds on the expected runtimes of probabilistic integer programs (\tool{Absynth} \cite{ngo2018BoundedExpectationsResource} and \tool{eco-imp} \cite{avanzini2020ModularCostAnalysis}).
As in the \emph{Termination Competition} \cite{TermComp}, we used a timeout of 5 minutes per example.
The first entry in every cell is the number of benchmarks for which the tool inferred the respective bound.
In brackets, we give the corresponding number when only regarding our new examples.
For example, \tool{KoAT\,\!+\,\!CFR} finds a finite expected runtime bound for 84 of the 90 examples.
A linear expected bound (i.e., in $\landau(n)$) is found for 56 of these 84 examples, where 12 of these benchmarks are from our new set.
$\mathrm{AVG(s)}$ is the average runtime in seconds on all benchmarks and $\mathrm{AVG^+(s)}$ is the average runtime on all successful runs.
\begin{table}[t]
	\makebox[\textwidth][c]{
		\begin{tabular}{l|cc|cc|cc|c|c|cc|c|c}
			                               & \multicolumn{2}{c|}{$\landau(1)$} & \multicolumn{2}{c|}{$\landau(n)$} & \multicolumn{2}{c|}{$\landau(n^2)$}
			                               & $\landau(n^{>2})$                 & $\landau(\mathit{EXP})$           & \multicolumn{2}{c|}{$< \omega$}     & $\mathrm{AVG^+(s)}$ & $\mathrm{AVG(s)}$                                           \\
			\hline \tool{KoAT\,\!+\,\!CFR} & 11                                & (2)                               & 56                                  & (12)                & 14                &     & 2 & 1 & 84 & (14) & 11.68 & 11.34 \\
			\hline \tool{KoAT}             & 9                                 &                                   & 41                                  & (1)                 & 16                & (1) & 2 & 1 & 69 & (2)  & 2.71  & 2.41  \\
			\hline \tool{Absynth}          & 7                                 &                                   & 35                                  &                     & 9                 &     & 0 & 0 & 51 &      & 2.86  & 37.48 \\
			\hline \tool{eco-imp}          & 8                                 &                                   & 35                                  &                     & 6                 &     & 0 & 0 & 49 &      & 0.34  & 68.02
		\end{tabular}
	}
	\vspace*{.4cm}
	\caption{Evaluation of CFR on Probabilistic Programs}
	\label{fig:evaluation}
	\vspace*{-.6cm}
\end{table}

The experiments show that similar to its benefits for non-probabilistic programs \cite{giesl2022ImprovingAutomaticComplexity}, CFR also increases the power of automated complexity analysis for probabilistic programs substantially, while the runtime of the analyzer may become longer since CFR increases the size of the program.
The experiments also indicate that a related CFR technique is not available in the other complexity analyzers.
Thus, we conjecture that other tools for complexity or termination analysis of PIPs would also benefit from the integration of our CFR technique.

\KoAT's source code, a binary, and a Docker image are available at:
\[\mbox{\url{https://koat.verify.rwth-aachen.de/prob_cfr}}\]
The website also explains how to use our CFR implementation separately (without the rest
of \KoAT), in order to access it as a black box by other tools.
Moreover, the website provides a \emph{web interface} to directly run \KoAT{} online, and details on our experiments, including our benchmark collection.

\paragraph{Acknowledgements:}
We thank Yoann Kehler for helping with the implementation of our CFR technique in \KoAT{}.

\printbibliography

\clearpage \appendix
\section{Formal Semantics of PIPs}
\label{app:pip}
For a detailed recapitulation of basic concepts from probability theory or the cylindrical construction of probability spaces for PIPs, we refer the reader to \cite{meyer2021InferringExpectedRuntimesReport}.
As mentioned in \cref{sec:preliminaries}, we use (history-independent) \emph{schedulers} to resolve non-deterministic branching and sampling.

\begin{definition}[Scheduler \cite{meyer2021InferringExpectedRuntimesReport}]
	\label{def:Scheduler}
	A function $\scheduler: \Conf \to \GTSet_\bot\times \State$ is a \emph{scheduler}
	if for every configuration $c = (\location,\state) \in\Conf$, $\scheduler (c) = (g,\tilde{\state})$ implies:
	\begin{enumerate}[label=(\alph*)]
		\item \label{scheduler:req1}
		      $\state (x) = \tilde{\state} (x)$ for all $x\in\PVSet$.
		\item \label{scheduler:req2}
		      $\location$ is the start location $\location_g$ of the general transition $g$.
		\item \label{scheduler:req3}
		      $\tilde{\state}\models\guard_g$ for the guard $\guard_g$ of the general transition $g$.
		\item \label{scheduler:req4}
		      $g = g_{\bot}$ and $\tilde{\state} = \state$, if $\location = \location_\bot$ or if no $g\in \GTSet, \tilde{\state}\in \State$ satisfy \cref{scheduler:req1,scheduler:req2,scheduler:req3}.
	\end{enumerate}
	$\Scheduler$ denotes the set of all schedulers.
\end{definition}
So to continue an evaluation in state $\state$, \cref{scheduler:req1} the scheduler chooses a state $\tilde{\state}$ that agrees with $\state$ on all program variables, but where the values for temporary variables are chosen non-deterministically.
After instantiating the temporary variables, \cref{scheduler:req2} the scheduler selects a general transition that starts in the current location $\location$ and \cref{scheduler:req3} whose guard is satisfied, if such a general transition exists.
Otherwise, \cref{scheduler:req4} $\scheduler$ chooses $g_{\bot} = \{ t_{\bot} \}$ and leaves the state $\state$ unchanged.

For a path $(\location,\state)$ of length $0$, i.e., if the path only consists of a single configuration, as mentioned in \cref{sec:preliminaries}, we have $$\prob{\scheduler}{\Program}(\location,\state) =
	\begin{cases}
		1 & \text{if $\location = \location_0$ and $\state = \initial$} \\
		0 & \, \text{otherwise.}
	\end{cases}
$$ Furthermore, for a transition $t$, two configurations $c = (\location_c,\state)$ and $c' = (\location_{c'},\state')$, and $\scheduler(\location_c,\sigma) = (g,\tilde{\sigma})$, we have $$\probTr{\scheduler}{\Program}(c\to_t c') =
	\begin{cases}
		p & \text{if $t = (\location_c, \guard, p, \update, \location_{c'}) \in g$, $t \neq t_\bot$,} \\
		  & \text{$\state'(v) = \tilde{\state}(\update(v))$ for all $v\in\PVSet$, and}                \\
		  & \text{$\state'(v) = \tilde{\state}(v)$ for all $v\in\VSet\setminus\PVSet$
		}                                                                                             \\
		1 & \text{if $t = t_\bot \in g$, $\location_{c'} = \location_\bot$, and $\sigma' = \sigma$
		}                                                                                             \\
		0 & \, \text{otherwise.}
	\end{cases}
$$ Then $\prob{\scheduler}{\Program}(c_0\to_{t_1}\dots\to_{t_n}c_n) = \prob{\scheduler}{\Program}(c_0)\cdot\prod_{i = 1}^n\probTr{\scheduler}{\Program}(c_{i-1}\to_{t_i} c_i)$.
For any $f \in \Path$, we say that $f$ is \emph{admissible} for $\scheduler$ and $\sigma_0$ if $\prob{\scheduler}{\Program}(f) > 0$.

\end{document}